# Visual Photometry: Testing Hypotheses Concerning Bias and Precision

**Alan B. Whiting**

*Department of Physics and Astronomy, University of Birmingham, Edgbaston, Birmingham BT2 5TT, UK; abw@star.sr.bham.ac.uk*

**Abstract**  Visual photometry, the estimation of stellar brightness by eye, continues to provide valuable data even in this highly-instrumented era. However, the eye-brain system functions differently from electronic sensors and its products can be expected to have different characteristics. Here I characterize some aspects of the visual data set by examining ten well-observed variable stars from the AAVSO International Database. The standard deviation around a best-fit curve ranges from 0.14 to 0.34 magnitude, smaller than most previous estimates. The difference in scatter between stars is significant, but does not correlate with such things as range or quickness of variation, or even with color. Naked-eye variables, which would be expected to be more difficult to observe accurately, in fact show the smallest scatter. The difference between observers (bias) is less important than each observer's internal precision. A given observer's precision is not set but varies from star to star for unknown reasons. I note some results relevant to other citizen science projects.

## 1. Introduction: visual photometry

It may seem surprising that nowadays, with precision electronic instruments widespread even among amateur astronomers, visual photometry is still widely practiced and useful. But the collective database of visual observations has unmatched coverage in time. Even the All-Sky Automated Survey (ASAS) returns rather sparse coverage of any particular variable star (Mayangsari *et al.* 2014), requiring sophisticated methods to recover details of a light curve. As an alternative, visual observations may be used to fill in the gaps (Holdsworth *et al.* 2013). In any case, the visual database allows a researcher access to much history; the 87-year span of Leibowitz and Formiggini (2015), for instance, could not be matched by electronic data. In addition, the initial discovery of unexpected behavior is often visual, as in Surina *et al.* (2014).

But visual photometry is different from the instrumental sort. In a sense there are no raw visual data: they are all heavily processed by the eye-brain system before even the observer is allowed access. As one obvious example, the generally logarithmic response of the eye (as opposed to the linear behavior of a CCD) has been carried over into the magnitude system we still use. There are other effects, from the well-substantiated, conveniently collected in the AAVSO *Manual for Visual Observing of Variable Stars* (AAVSO 2013), to the anecdotal. Thus the data must be handled differently, and assumptions about their behavior can be dangerous. As an example, Pierce and Jacoby (1995) used visual data on a historical supernova in a determination of Hubble's constant; Schaefer (1996) came to a different conclusion based on a model of visual response. In reply, Jacoby and Pierce (1996) disagreed with Schaefer's method. The point is that a model and an analysis of data gave different results.

My present aim is to work out some characteristics of the visual database, taking mostly a consumer's viewpoint. The actual practice of visual photometry is relevant only as far as it suggests hypotheses to be tested. In these hypotheses I do not claim to be complete; many more aspects of the data remain to be investigated. A referee has suggested age and experience of the observer (unfortunately, difficult to test with the publicly available AAVSO data) as well as possible variation over time periods of years. No doubt others will occur to the reader as we proceed. Whiting (2012) has already considered some aspects of comparison stars.

Previous work (e.g. Stanton 1999; Collins 1999; Zissell 2003) has generally dealt with the color term, that is, on how to transform visual observations to a standard instrumental filter. Here I concentrate on internal statistics, leaving a connection to instrumental data for later studies. It seems best to have a more detailed and reliable picture of visual data before comparing them to other forms. And of course if the spread of visual data is whole magnitudes, as some authors report (Williams 1987; Price *et al.* 2007), a small color term is hardly worth applying.

In this study I look at visual data on ten well-observed variable stars from the AAVSO International Database of the American Association of Variable Star Observers (AAVSO), calculating their residuals around the best-fitting curve. My aim is to determine the size of these residuals and work out what factors affect that size. Ten is of course an inadequate sample on which to base conclusions about tens of thousands of variable *stars*. However, thousands of observations by hundreds of observers constitute quite a firm foundation for conclusions about the *data*.

One important theme is comparing differences between observers with an observer's internal variation. Here, I will use "bias" to mean the average difference between an observer's data and the best-fitting curve; "precision" to mean an observer's standard deviation about that average; and "scatter" or "accuracy" to mean the combined standard deviation about the curve.

## 2. Data selection and processing

### 2.1. The stars

We need stars with many observations, not only to produce a well-defined light curve but to populate the residuals around it. Beyond that, we would like stars with different characteristics in order to investigate possible effects of the type of variation. The final sample includes three Mira-type long period variables, five semiregular variables (three of them naked-eye stars) and two carbon stars. The selection is not intended to be exhaustive or representative, but to test certain *a priori* plausible effects (detailed below). The data on TX Piscium were sparser than those of the other stars, which probably had a minor effect on its results.

The premier source of visual photometric data for researchers is the on-line portal of the AAVSO, used for all the data in this study (Kafka 2017). Unfortunately it is not practical to list the thousands of data points individually. For each star I limited the data to a single full apparition, to avoid problems with curve-fitting over a gap. All the data were downloaded from the AAVSO web site, using only those points identified as visual. In what follows I use "days" to mean Julian Day minus 2457000.

Points identified as "fainter than" were not used. Data flagged as "magnitude uncertain" were included, since they are part of the database and indeed might have told something about it. Unfortunately, there were not enough of them (12 out of a total of 8091) to allow much of a conclusion. No other selection was performed, since the aim was to characterize the data, not to study the stars.

2.2. Fitting functions

Getting the fit right in detail is more important for this study than for other types of analysis. A curve that places the maxima and minima at the right places and times, for instance, would be sufficient for determining the period and amplitude of a Cepheid or RR Lyrae. However, if it followed the rising branch of Mira's curve (see Figure 1) with too steep or too shallow a slope, it would give systematic offsets that would change the shape of the residual dispersion and possibly throw off the answers.

A simple smoothing is the common way to deal with visual data, but would tend to flatten extrema and thus possibly distort the residuals. I tried Legendre polynomials, but as terms were added artifacts appeared (sections that obviously departed from the trend of the observations) before a good fit was obtained. The periodic Gaussian functions of Inno *et al.* (2015) looked promising, but I was unable to get them to converge on these data. In the end I fell back on a Fourier expansion plus a linear term. I used the IDL "curvefit" routine, which performed a least-squares fit.

2.3. The fit

For each star, I started with a few Fourier modes and added terms until a decent fit by eye was obtained. Initially I hoped to be able to use the residuals about each fit to decide when to stop. The process of fitting Mira is shown in Table 1. The standard deviation of the residuals around the fit is shown. Their Gaussian character was tested by running a $\chi^2$ comparison (which gives the probability of two distributions being the same, within expected fluctuations) with the normal distribution of the same mean and standard deviation, also computing skewness and kurtosis.

As one might expect, a fourth-order fit gets the gist of the variation but doesn't follow it closely. Fifth- and sixth-order fits don't fit the rising branch very well. The seventh-order fit captures the rising branch, but has a double minimum that I reject as unphysical; only by going to tenth-order do we get a

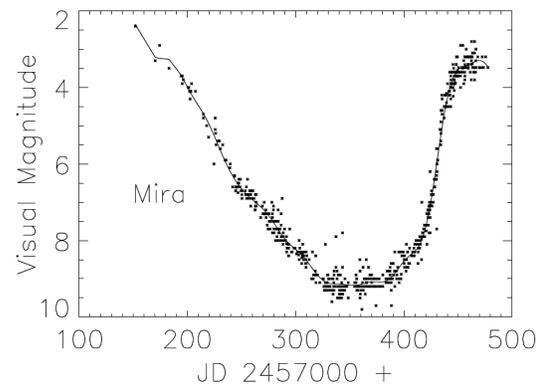

Figure 1. Visual observations of Mira, o Ceti, with the best-fitting average light curve superimposed. The most challenging task with fitting a curve proved to be following the steep rising branch without adding artifacts.

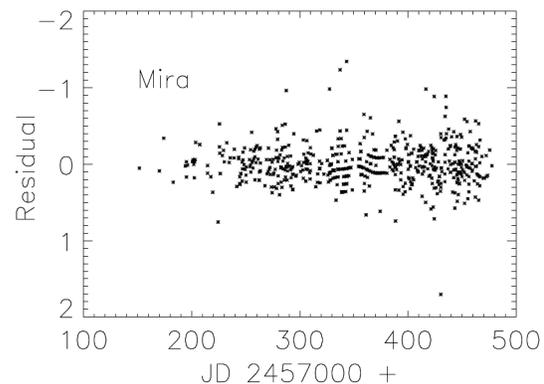

Figure 2. Residuals of the observations of Mira from the fitted light curve. The practice of most (but not all) visual observers of reporting to the nearest tenth-magnitude leads to some artifacts.

Table 1. Statistics for the best Fourier fit to the Mira visual observations.

| Order | σ | $P(\chi^2)$ | Skew | Kurtosis |
|---|---|---|---|---|
| 4 | 0.329 | 0.0007 | –0.129 | 2.47 |
| 5 | 0.298 | 0.0 | –0.402 | 4.02 |
| 6 | 0.283 | $6 \times 10^{-6}$ | –0.293 | 4.27 |
| 7 | 0.268 | $1.8 \times 10^{-7}$ | –0.309 | 5.12 |
| 8 | 0.262 | $6.0 \times 10^{-8}$ | –0.323 | 5.67 |
| 9 | 0.258 | $1.1 \times 10^{-6}$ | –0.367 | 6.00 |
| 10 | 0.256 | $6.0 \times 10^{-8}$ | –0.345 | 6.10 |

*Note: From left to right: order of fit, scatter about the fit, probability of matching a normal distribution based on $\chi^2$ (less than $10^{-8}$ shown as 0), skew, kurtosis.*

single minimum. The data and the fit are shown in Figure 1, and residuals (data minus the fit) in Figure 2.

Adding another Fourier term will *always* give a smaller scatter. The question here, as in many other situations, is whether the added term does any good; that is, does it reduce the scatter enough to make it worthwhile? For this we employ the F-ratio test on the variances, which gives the probability of the added term being useful.

Applying an F-ratio test on the Mira fits, all higher orders are significantly better fits than fifth (at the one-percent level), but order-by-order there is no preference beyond sixth and the tenth is no better than the seventh at the five-percent level. Thus, overall statistics of the residuals are no help at fitting the light curve. On the other hand (and this is important), the statistics are insensitive to details of the fit. Even a curve with obvious artifacts gives essentially the same statistics. (The F-ratio test assumes Gaussian behavior in the variance, which is not strictly true here. However, the qualitative conclusion stands: the scatter is not useful as a guide to the quality of fit.)

We want a better criterion than a by-eye fit and will need one for lower amplitude variables. To this end, consider the situation in Figure 3. Here the fit at a given order is shown by the smoother curve, while adding the next gives oscillations of a certain amplitude and wavelength around it. Focusing on one of these oscillations, evidently the average of the observations is displaced from that of the smoother curve by a certain amount. We ask: what is the probability that this happens by a chance fluctuation of residuals? The probability will be higher if there are fewer data points, if the amplitude is small, and if the standard deviation of the curve is large. Conversely, if the data points are dense and fit tightly around the curve, we will be able to detect a smaller real amplitude.

If the residuals are Gaussian with standard deviation σ, the chance of an offset by Δy over a number of observations n is related to Student's t-statistic:

$$t = \frac{\Delta y}{\sigma}\sqrt{n}. \qquad (1)$$

We proceed as follows: from a previous curve of standard deviation σ, require any fluctuation to have a probability by chance of 5% or less over a number of observations n = N/m, where N is the total for the star and m the Fourier order. Using Equation 1 we find a threshold Δy, above which we accept the higher-order fit and below which we reject it.

This procedure should not be regarded as fully rigorous and quantitative. As noted in Table 1, the residuals are not Gaussian (this is true throughout our sample). More importantly, the amplitude of a high-order fit depends on observations outside a single oscillation (as we will see below). However, it does provide a consistent criterion for terminating the fitting procedure, and does answer the requirement that a higher density of observations is needed in order to accept higher-frequency and lower-amplitude features of the light curve (as noted by Trumpler and Weaver (1953)).

A summary of the input observations is provided in Table 2.

R Andromedae, o Ceti (Mira), and R Leonis are Mira-type long period variables. U Monocerotis and R Scuti are large-amplitude semiregular variables, while α Herculis, α Orionis and μ Cephei are smaller-amplitude, naked-eye semiregular types. TX Piscium and V Aquilae are carbon stars.

2.4. The Full Moon effect

In the tenth-order fit for Mira there remains a stubborn oscillation of about 0.03 mag amplitude with a thirty-day period. Using the t-statistic criterion I reject it as an artifact of the data; but efforts to smooth it out proved unavailing. Looking more closely at one section of the curve, near the minimum (see Figure 4), we find that there is a thirty-day oscillation in the *number* of observations. About day 320, 350, and 380 there are few, with pulses of activity between. Following a hunch, I found that days 322, 352, and 382 were Full Moons. Clearly the

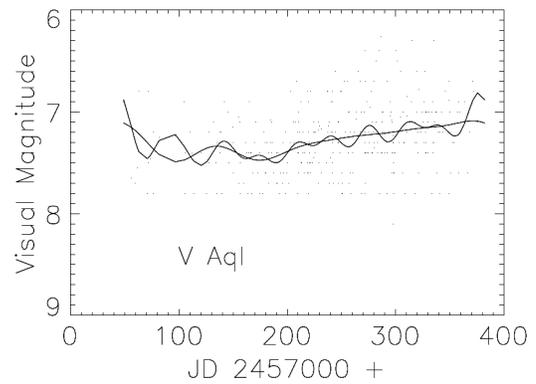

Figure 3. A step in the fitting of the light curve to V Aquilae. A lower-order fit gives the smoother curve; adding an additional order gives the more oscillatory one. Using the criterion developed in the text, the higher-order correction is rejected as noise.

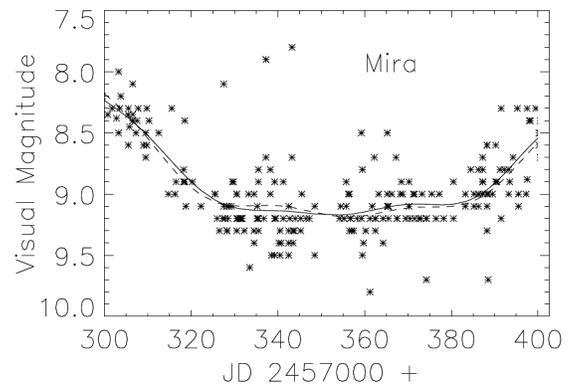

Figure 4. Observations of Mira along with the tenth-order fit (smooth curve) in the region of the minimum. Note the low-amplitude thirty-day oscillation.

Table 2. Summary of the input visual observations of ten variable stars.

| Star | Obns | Obsrs | Obsn day$^{-1}$ | Order |
|---|---|---|---|---|
| α Herculis | 634 | 35 | 1.94 | 5 |
| α Orionis | 603 | 72 | 2.05 | 6 |
| μ Cephei | 1754 | 105 | 4.05 | 4 |
| U Monocerotis | 781 | 70 | 2.67 | 10 |
| R Andromedae | 386 | 62 | 0.99 | 7 |
| R Scuti | 2118 | 148 | 6.08 | 15 |
| TX Piscium | 124 | 13 | 0.45 | 2 |
| o Ceti | 594 | 88 | 1.82 | 10 |
| R Leonis | 745 | 124 | 2.74 | 6 |
| V Aquilae | 352 | 47 | 1.06 | 3 |

*Note: The columns are: star designation; number of observations; number of observers; average observations per day; order of best Fourier fit.*

amateur astronomers who followed Mira were also looking at other deep-sky objects, and preferred dark skies for their work! This pulsing of observations apparently injected a signal into the thirty-day Fourier mode.

Such a Full Moon effect is also visible in the other Miras in our sample, R Leonis and R Andromedae, though not in the other stars. The amplitude is small, and it has no effect on the overall statistics. But the implications for this and other citizen-science efforts are large: unexpected artifacts, statistically robust, can appear without warning.

 

## 3. Results and discussion

3.1. Residuals

A summary of the fits and residuals is provided in Table 3.

The first thing to notice is the size of the residuals ("Scatter" in the table). Previous work has generally given much larger figures. Williams (1987) determined a value of 0.5 magnitude from variations in field orientation alone, though that was for a single observer. Simonsen (2004) cites 1.5–2.0 mag, without giving details. Price *et al*. (2007) show a scatter ranging from 0.2 to 1.0 mag, heavily dependent on spectral type. In contrast, Whiting (2012) gives a scatter of 0.2 to 0.3 mag for visual observations of Miras.

The study of Price *et al*. (2007) is, however, problematic. They took data from the AAVSO database on 3542 stars, each with 1000 or more observations, and subtracted a tenth-order polynomial. They do not say what time span was covered, though they imply that multiple periods were included. There is no indication of how well the polynomial fit the actual light curve, or indeed that they considered the matter. It is impossible to tell what part of their results arise from a poor fit and what part is real. Their work is, therefore, not useful for the present study.

What about the data points flagged as "magnitude uncertain?" Mira has one such point, with a residual of –0.526, twice the size of the overall scatter, though not the largest residual for that star. R Andromedae has two, which at –0.143 and 0.243 are unremarkable. The well-observed μ Cephei has eight, whose standard deviation of 0.249 is greater than that of the whole dataset, though not by much. The single "uncertain" data point for V Aquilae has a residual of –0.013, much *smaller* than the run of the data. No other stars have data so flagged. From this small sample all we can conclude is that there is no reason to exclude "magnitude uncertain" data from any study, and they are included here.

The next thing to notice is that the scatter varies appreciably from star to star. There is, thus, no single figure for "the accuracy of visual photometry." Reference to Table 2 shows that it's not a matter of number or density of observations, or the Fourier order of the fit. Where else might it come from?

People are known to see what they expect to see. Perhaps a star varying swiftly, or over a great range, would be harder to follow. To test this, each star's speed of variation as well as its total range are listed in Table 3. For the speed, the slope of the fitting function was evaluated at each data point, and the standard deviation of these slopes calculated. Neither range nor speed shows a correlation with scatter, as is shown by Figures 5 and 6. Nor is predictability important: the semiregular variables show no overall tendency toward larger residuals than the more predictable Miras.

Surely color will play a part, since color perception varies widely among people. Indeed, V Aquilae, one of the reddest stars known, has the largest residuals in the sample. However, TX Piscium, the next reddest star, shows no unusual scatter. Or consider pairs of stars: TX Piscium and o Ceti differ in color more than do Betelgeuse and Rigel, and they have the same scatter. R Leonis and V Aquilae differ by even more, and again have indistinguishable accuracy. If there were a color effect it

Table 3. Summary of the fits and residuals for the ten variable stars.

| Star | Scatter | B–V | dm/dt | δ m | Skew | Kurt |
|---|---|---|---|---|---|---|
| α Her | 0.141 | 1.45 | 3.8 | 0.39 | 1.78 | 5.53 |
| α Ori | 0.197 | 1.85 | 3.5 | 0.37 | 0.020 | 0.019 |
| μ Ceph | 0.212 | 1.35 | 2.2 | 0.40 | 0.512 | 0.895 |
| U Mon | 0.227 | 1.18 | 34.1 | 1.12 | 0.328 | 0.339 |
| R And | 0.245 | 1.97 | 55.8 | 8.32 | –2.60 | 1.56 |
| R Sct | 0.249 | 1.47 | 46.4 | 1.80 | 0.053 | 1.12 |
| TX Psc | 0.252 | 2.60 | 3.8 | 0.58 | –0.334 | 0.278 |
| o Cet | 0.256 | 1.10 | 63.1 | 6.82 | –0.345 | 6.11 |
| R Leo | 0.329 | 1.41 | 31.0 | 4.91 | –0.503 | 1.29 |
| V Aql | 0.337 | 4.19 | 1.7 | 0.50 | –0.004 | 2.15 |

*Note: The columns are: star designation; standard deviation of the residuals, in mag; B–V color; standard deviation of the speed of variation, in mmag day$^{-1}$; total amplitude of variation (minimum to maximum), mag; skew; excess kurtosis.*

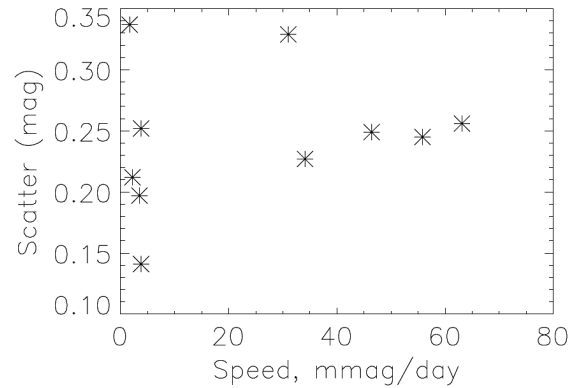

Figure 5. Scatter of the residuals of the ten stars plotted against the speed of variation (standard deviation of the slope of the fitting function, as evaluated at each data point). No correlation is evident.

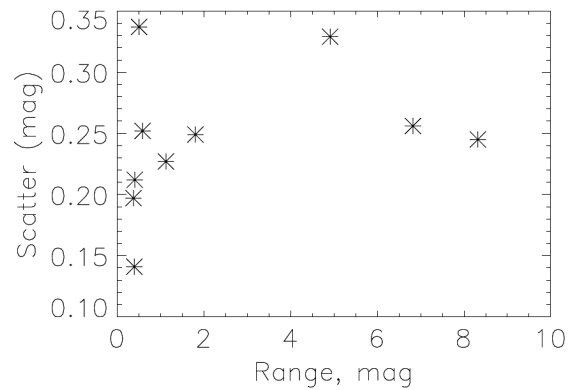

Figure 6. Scatter of the residuals of the ten stars plotted against the total range of variation. No correlation is evident.

would certainly show up in these pairs, and it doesn't. A color-scatter plot is shown in Figure 7.

We come to the surprising conclusion that the color of a star does not seem to affect the accuracy of its brightness estimate. This is not unprecedented, however; Whiting (2012) found that the colors of comparison stars had no effect on the scatter of visual estimates of Mira variables.

One might guess that naked-eye stars would be more difficult subjects, since comparison stars will generally be

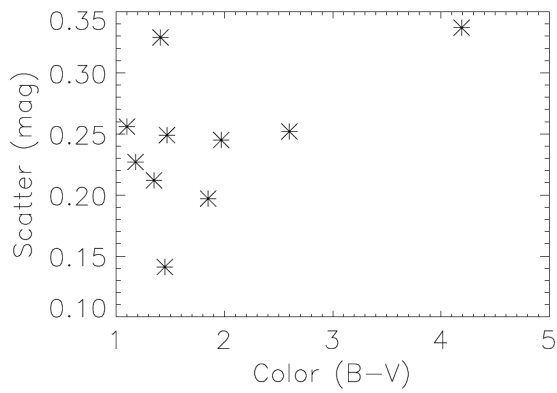

Figure 7. Scatter of the residuals of the ten stars plotted against B–V color. Note that the contrasting stars of Orion, Rigel and Betelgeuse, have a difference in B–V color of about 1.48, much less than is covered here.

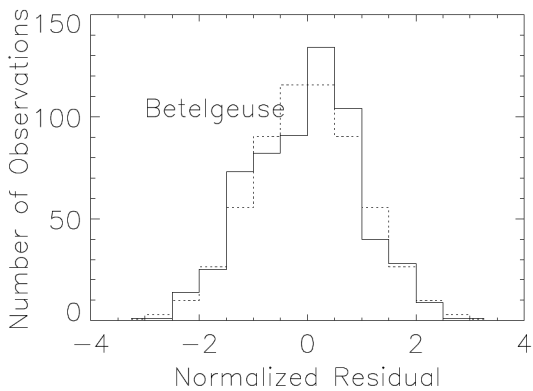

Figure 8. Residuals of observations of Betelgeuse (solid histogram) compared to a Gaussian of the same standard deviation (dotted histogram). The skew is evident. This star has the lowest excess kurtosis of the sample.

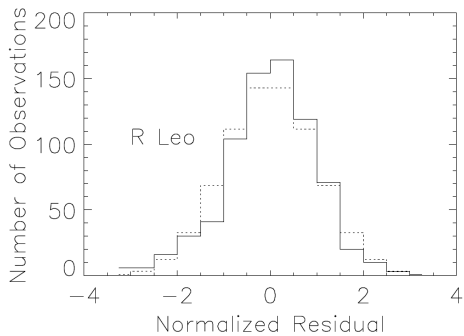

Figure 9. Residuals of observations of R Leonis (solid histogram) compared to a Gaussian of the same standard deviation (dotted histogram). Here the excess kurtosis is clear.

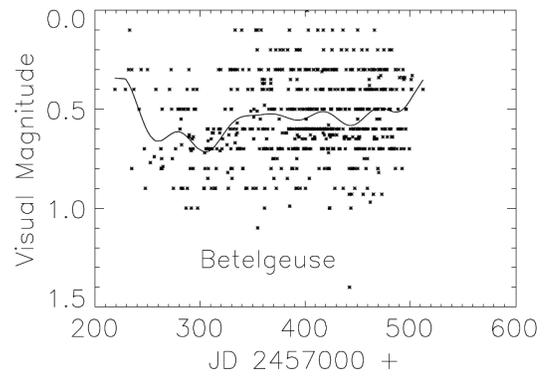

Figure 10. Visual observations of Betelgeuse, α Orionis, with the fitted light curve.

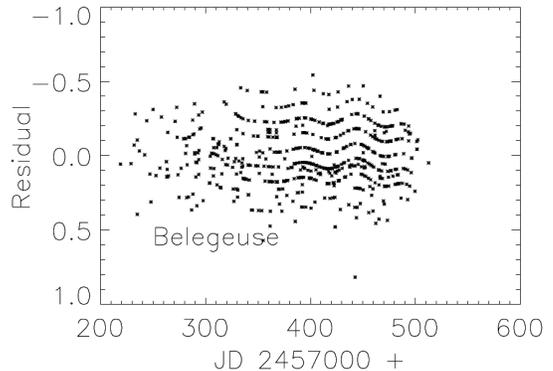

Figure 11. Residuals of the observations of Betelgeuse from the fitted light curve. The practice of most (but not all) visual observers of reporting to the nearest tenth-magnitude leads to some artifacts.

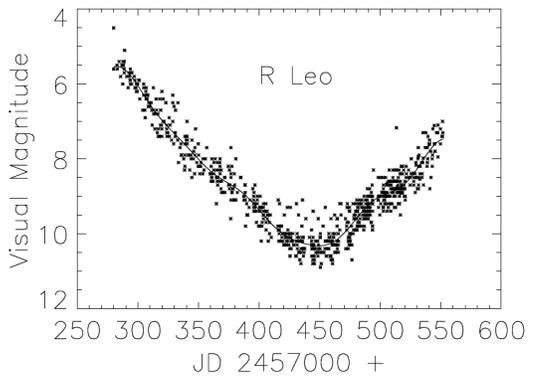

Figure 12. Visual observations of R Leonis with the fitted light curve.

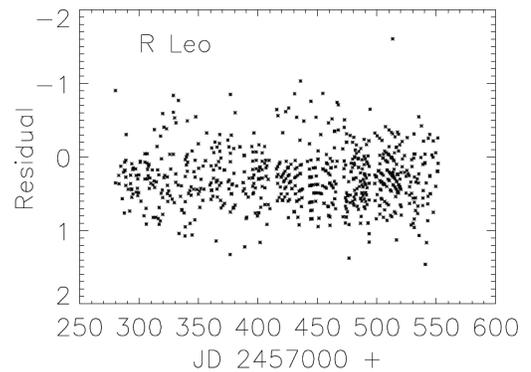

Figure 13. Residuals of the observations of R Leonis from the fitted light curve. The practice of most (but not all) visual observers of reporting to the nearest tenth-magnitude leads to some artifacts.

farther away in angular distance (and thus harder to keep in sight at the same time), and one is more likely to be distracted by other lights, to say nothing of airmass corrections. But the three naked-eye variables show the *least* scatter.

Thus, while the accuracy of visual photometry varies significantly among the stars in this sample, several plausible reasons for it do not apply.

One further feature of our sample brought out in Table 3 is the shape of the residuals. In all stars these are significantly non-Gaussian. For the most part they are not drastically different, but the departure is visible when plotted. (The exception is TX Piscium; for this star, I suggest that the observations are simply too few and sparse to rule out normality, rather than that they obey a different distribution). They all agree in having excess kurtosis (a sharper peak and more outliers than Gaussian) and significant skew, though there is no agreement on the sign of the latter. Examples of the shape of residuals compared to Gaussian are shown in Figures 8 and 9.

At this point our analysis proceeds by making eight plots for each star. It is obviously impossible to present all 80 of these within the confines of a paper. Instead, I include representatives of each type, generally R Leonis and Betelgeuse as high and low scatter stars, respectively, or others that illustrate specific results. All the plots are available in Appendix A. Fits and residuals for Betelgeuse and R Leonis are shown in Figures 10, 11, 12, and 13.

Visual observations are reported to the nearest tenth of a magnitude, a quantization obvious in the low-amplitude plots. Could this be the source of the non-Normal shape? To test this, I took the fitted curve, added random Gaussian noise, rounded to the nearest tenth-magnitude, and re-ran the fitting procedure. In each case the Gaussian character (and standard deviation) of the synthetic residuals was returned. Quantization affects neither the size nor the shape of visual residuals.

However, the re-fitted curves are not exactly the same as the originals. For most of the stars and most of the time the difference is of the order 0.03–0.04 magnitude. To see this it is necessary to zoom in on a small section of the data, for for instance as in Figure 4, where the smooth curve is the original fit and the dashed curve the re-fit. However, in U Monocerotis local minima sometimes fail by a larger amount (see Figure 14). In V Aquilae, sparse observations during the beginning and end of the apparition create greater uncertainty there (Figure 15). It is beyond the scope of this paper to work out rigorously how firmly the curves are determined by the data, but these examples should give some idea.

3.2. Demographics

How much does it matter who does the observing? In particular, can the difference in residuals between stars be explained by populations of observers with greater or lesser accuracy?

As a first step in studying the demographics of visual photometry, I break down the observers by number of observations. That is, I count the number of observers who submitted one observation of a particular star, two observations, and so forth. Two representative histograms are shown in Figures 16 and 17.

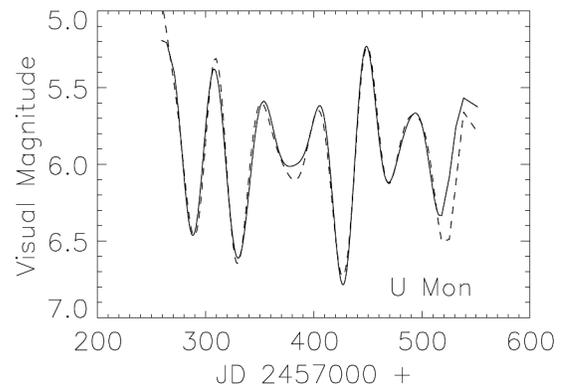

Figure 14. Original fitted light curve of U Monocerotis (solid curve) compared with one with synthetic Gaussian residuals (dashed curve). They fail to match the amplitudes in some minima.

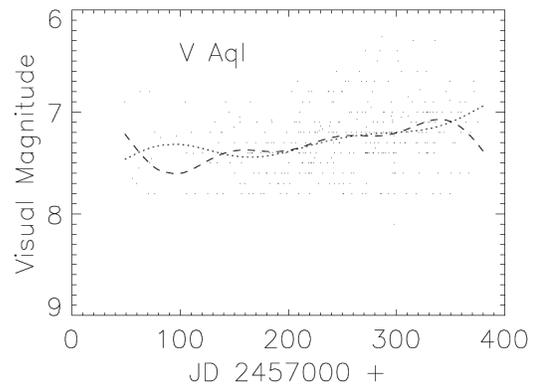

Figure 15. Original fitted light curve of V Aquilae (dashed curve) compared with one with synthetic Gaussian residuals (dotted curve). They match relatively poorly during the sparsely-observed beginning and end of the apparition.

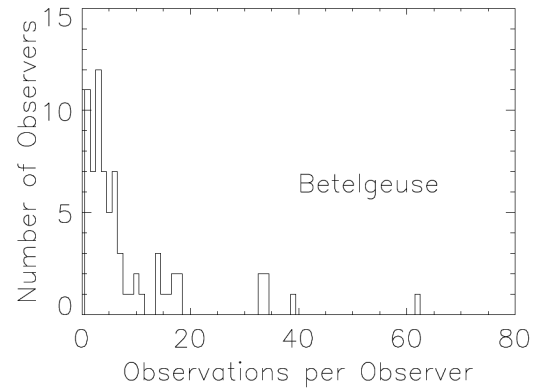

Figure 16. Number of observers of Betelgeuse compared to the number of observations each submitted. The database is dominated neither by the few-observation observers nor the handful of very active ones.

There are more observers with few observations than with many. However, counting up the contributions the mass of *observations* is not dominated by either end of the spectrum. The same holds true for all our stars (with the exception of TX Piscium, which has a relative lack of low-activity observers). The difference in residuals between stars thus cannot be attributed to observer populations of higher or lower activity. If we make the plausible assumption that low-activity observers are those with less experience, the difference is not due to some stars being popular with neophytes and others with veterans. (This assumption is far from certain. There are many reasons why an experienced observer might only submit a few estimates for a given star in a given season. However, it is probably true overall, and one whose observations are in the dozens certainly has gained some experience.)

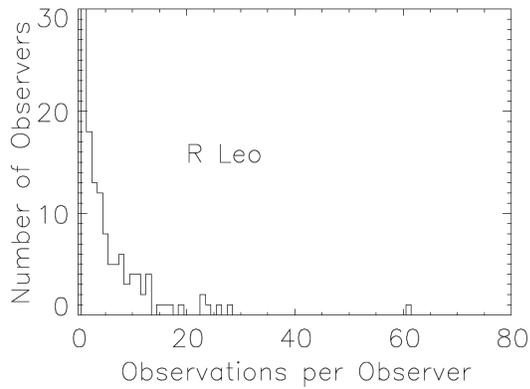

Figure 17. Number of observers versus number of observations per observer for R Leonis. Again, the mass of observations is not dominated by either end of the spectrum of activity.

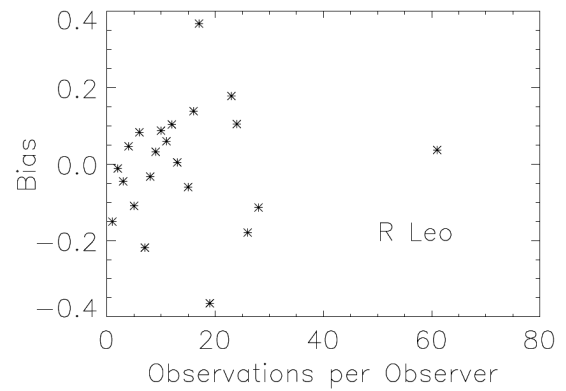

Figure 20. Bias (average residual about the fitted curve) for observations of R Leonis, broken down by number of observations made by the observer.

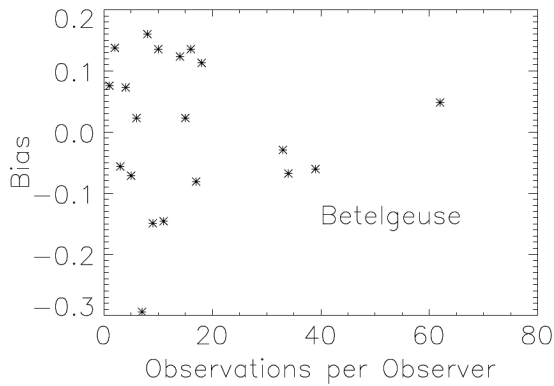

Figure 18. Bias (average residual about the fitted curve) for observations of Betelgeuse, broken down by number of observations made by the observer.

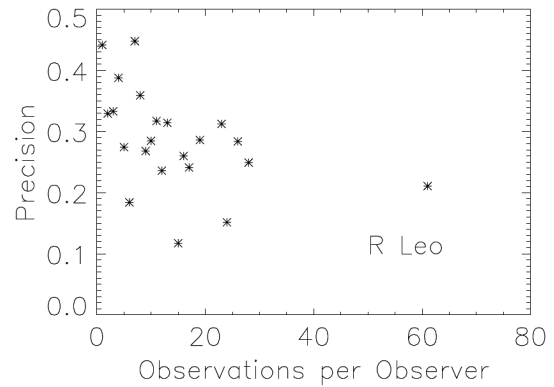

Figure 21. Precision (standard deviation of residuals) for observations of R Leonis, broken down by number of observations made by the observer.

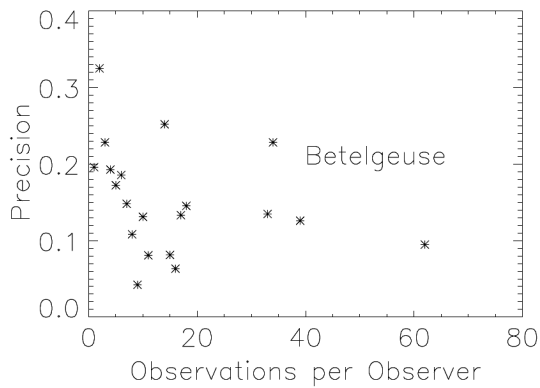

Figure 19. Precision (standard deviation of residuals) for observations of Betelgeuse, broken down by number of observations made by the observer.

Pursuing the question of observer populations further, I break down the residuals for each observer into the average (bias) and the standard deviation about that average (precision). Representative plots for bias and precision as a function of number of observations are shown in Figures 18, 19, 20, and 21.

In each case, out to about thirty observations there is no apparent advantage to additional experience. It might appear that the points beyond are more precise or of smaller bias, but there are really too few to conclude that.

The next step is to compare the relative size of bias and precision. Plotting them against each other for each observer, something like Figures 22 and 23 results (corresponding plots appear in Appendix A). Straight lines are included to show where bias and precision are equal in magnitude, and error bars on bias produced by dividing the figure by the square root of the number of observations. (Error bars are not included for precision to avoid excessive clutter.)

There are observers more or less evenly spread over the plots, which would indicate that bias and precision contribute roughly equal amounts to residuals. For those observers inside the "funnel," bias is less important; outside, bias is more important. However, observers outside the funnel are predominantly those with large error bars and hence few observations. Most of the *observations* are inside the funnel, showing that observer-to-observer bias contributes less to the residuals than the precision of each observer's data.

A summary of the combined bias and precision for each star is given in Table 4. As shown by Figures 22 and 23, the contribution of bias to total scatter is somewhat overstated

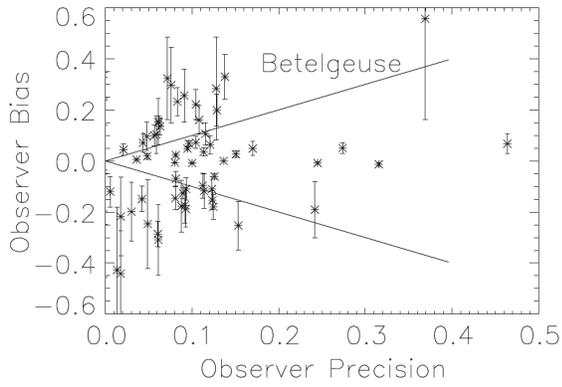

Figure 22. Bias (average of residual around the fitted curve) compared with precision (standard deviation around the average) for the observers of Betelgeuse. The straight lines show where the quantities are equal *in magnitude*. Error bars are produced by dividing the bias by the square root of the number of observations; similar bars for precision are omitted for clarity.

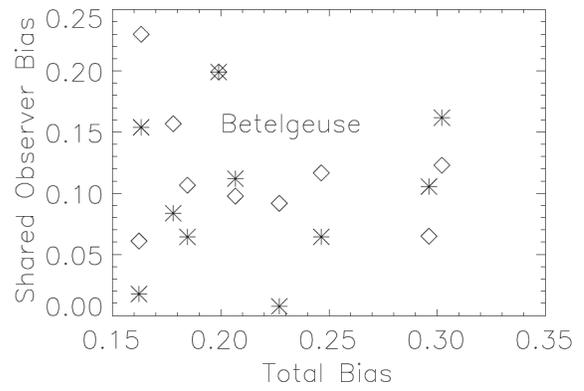

Figure 25. Comparison of the bias of sets of observers of Betelgeuse and another star. For each star, the standard deviation of the bias of the shared observers at Betelgeuse is shown by a diamond, at the other star by an asterisk; the ordinate is the standard deviation of the bias of all observers of the other star. Betelgeuse observers show higher bias at home than at low-bias stars, lower at home than high-bias stars.

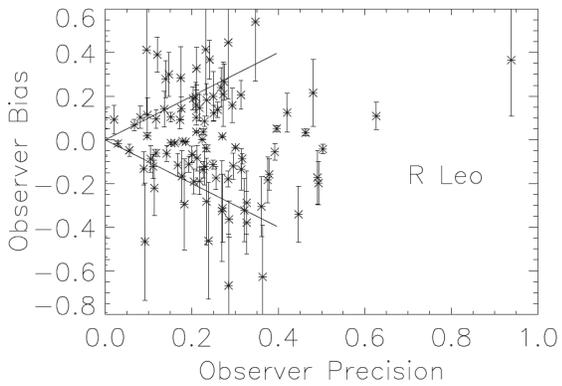

Figure 23. Bias (average of residual around the fitted curve) compared with precision (standard deviation around the average) for the observers of R Leonis. The straight lines show where the quantities are equal *in magnitude*. Error bars are produced by dividing the bias by the square root of the number of observations; similar bars for spread are omitted for clarity.

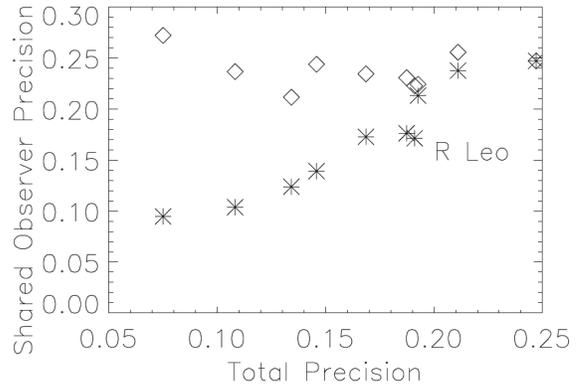

Figure 26. Comparison of the precision of sets of observers of R Leonis and another star. For each star, the average of the precision of the shared observers at R Leonis is shown by a diamond, at the other star by an asterisk; the ordinate is the average precision of all observers of the other star. The shared observers have indistinguishable precision at R Leonis, but their performance at other stars marches in step with that of all observers of that star.

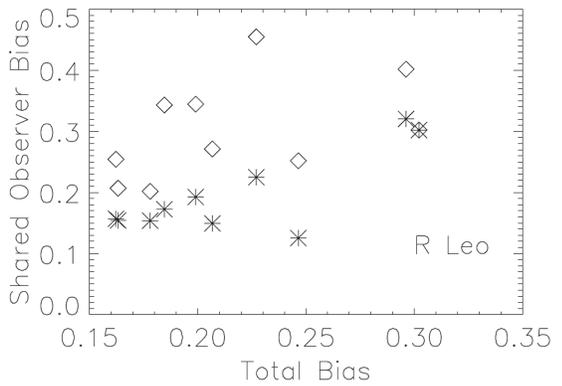

Figure 24. Comparison of the bias of sets of observers of R Leonis and another star. For each star, the standard deviation of the bias of the shared observers at R Leonis is shown by a diamond, at the other star by an asterisk; the ordinate is the standard deviation of the bias of all observers of the other star. The consistent trend is for observers to be less accurate at R Leo itself.

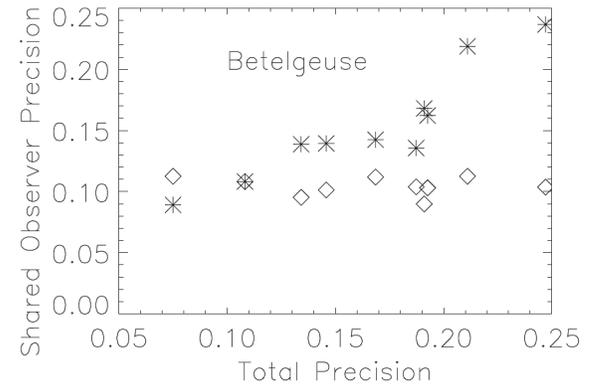

Figure 27. Comparison of the precision of sets of observers of Betelgeuse and another star. For each star, the average of the precision of the shared observers at Betelgeuse is shown by a diamond, at the other star by an asterisk; the ordinate is the average precision of all observers of the other star. The shared observers have indistinguishable precision at Betelgeuse, but their performance at other stars marches in step with that of all observers of that star.

Table 4. Contribution of bias and precision to the total scatter of the residuals of visual observations.

| Star | B−V | Bias | Precision | Scatter |
|---|---|---|---|---|
| α Her | 1.45 | 0.163 | 0.075 | 0.141 |
| α Ori | 1.85 | 0.199 | 0.108 | 0.197 |
| μ Ceph | 1.35 | 0.207 | 0.146 | 0.212 |
| U Mon | 1.18 | 0.162 | 0.169 | 0.227 |
| R And | 1.97 | 0.246 | 0.187 | 0.245 |
| R Sct | 1.47 | 0.184 | 0.211 | 0.249 |
| TX Psc | 2.60 | 0.227 | 0.134 | 0.252 |
| o Ceti | 1.10 | 0.178 | 0.192 | 0.256 |
| R Leo | 1.41 | 0.302 | 0.247 | 0.329 |
| V Aql | 4.19 | 0.296 | 0.191 | 0.337 |

*Note:* "Bias" is the standard deviation of the observers' biases; "Precision" is the mean of the observers' precisions; "Scatter" is the standard deviation of all the residuals of a star. (They do not add in quadrature due to the varying number of observations per observer.) The second and fifth columns are repeated from Table 3 for convenience.

by the average figure given. Note that bias for R Leonis, a star of no unusual color, is the same size as for V Aquilae, an extremely red star. This underlines the fact that a difference in color perception between observers is not an important source of scatter.

Although we have concluded that accuracy is not correlated with the activity of a given observer, the possibility remains that stars with a smaller scatter owe it to observers with smaller residuals. Indeed, there is anecdotal evidence for elite observers accurate to 0.05 mag. Under this hypothesis, the small scatter of Betelgeuse is due to a group of intrinsically more accurate observers, while the large scatter of R Leonis is due to a less accurate group. To test this, we look at observers who submitted estimates on more than one star. The nominally elite Betelgeusans should perform as well on other targets, while the Leonids should have a consistently large scatter. (The observers of, say, both R Leonis and R Scuti would not necessarily be identical with those of R Leonis and μ Cephei, so we need to look at each group's performance at each star.)

First we plot the standard deviation of their bias at the "home" star with diamonds, then at the "away" star with asterisks, using as an ordinate the standard deviation of all observers' bias at the "away" star. The resulting plots are Figures 24 and 25. It is immediately apparent that there is no consistent "observer bias" even for limited subsets of observers; the figure can vary by a factor of two or more from star to star. R Leonis observers consistently have a smaller bias at the "away" star; Betelgeuse observers are less systematic, though there is a tendency to have a larger bias at Betelgeuse for stars with smaller total bias, and a smaller bias at Betelgeuse for stars with a larger total bias. Looking only at "away" stars, R Leonis observers tend to have a larger average bias than Betelgeuse observers, but there is much overlap.

Continuing the investigation with observer precision we obtain Figures 26 and 27. The shared observers' performance at their home star is very consistent: we have not picked out unusual sets of observers. Their spread at the "away" star marches in step with that of all other observers. Indeed, comparing the performance of nominally high-precision Betelgeusans with that of nominally low-precision Leonids, we find them essentially the same. We conclude that the variation of precision is in our stars, not in ourselves.

## 4. Conclusions

Some of the results reported here should be encouraging both to the users and producers of visual photometry. The internal accuracy of the method is tighter than several previous works have reported. Bias, the difference between observers, is less important than precision, the spread of each observer's residuals; moreover, the precision of any set of observers at a particular star seems to be about the same. Low-activity observers have accuracy similar to more productive ones. It is unlikely, therefore, that the light curve of any particular star will suffer from bias or inaccuracy through an unlucky choice of observers.

If we identify low-activity observers with newcomers, they should be encouraged that even their first observations are useful. This is in contrast with, for example, another citizen science project, the Galaxy Zoo. As shown by Figure 2 of Willett *et al.* (2013), users who classified fewer than 100 galaxies had low scores for consistency, and consistency increased with activity up to 1000 galaxies. I suggest that the usefulness of data from new variable star observers comes from the fact that the task is simple (which is *not* the same as easy), compared with the several steps of classifying galaxy images. (Probably the extreme visual task in astronomy is measuring double stars, where a year of steady work is necessary before producing any useful data at all (Argyle 2009); for full competence, Couteau (1981) desires eight or nine years, with a year's delay if switching to another telescope.) Veteran observers should be encouraged that their work on difficult objects (like carbon stars) is, in general, no less accurate than on apparently easier targets.

On the other hand, the present work has thrown up several puzzles. Visual accuracy varies from star to star with no obvious pattern; several plausible explanations fail to fit the data. The non-Gaussian character of residuals also awaits explanation. I note in passing that the existence of two (not several) populations of observers with different means and standard deviations might produce something like this, but that is only speculation.

It is possible that the small number of subjects in this study, ten stars of three different types, have somehow biased the results. In principle, the scatter around a cataclysmic variable curve or that of a supernova could look different. But it is very hard to see how. With 8091 data points from 319 observers, the characteristics of visual photometry seem well-established.

The curve-fitting procedure adopted here is adequate for the purpose, but could be improved, in particular to eliminate sensitivity to periodic noise.

For other projects involving citizen science, note that features of the data that are expected to be present (like greater uncertainty for redder stars) might not actually be there, while unexpected effects (as from the Full Moon) can appear. Even obvious things may need checking. That is perhaps the overall lesson, when people are concerned.

## 5. Acknowledgements

## References

AAVSO. 2013, *AAVSO Manual for Visual Observing of Variable Stars*, AAVSO, Cambridge, MA.

Argyle, R. 2009, personal communication.

Collins, P. L. 1999, *J. Amer. Assoc. Var. Star Obs.*, **27**, 65.

Couteau, P. 1981, *Observing Visual Double Stars*, MIT Press, Cambridge, MA, 86.

Holdsworth, D. L., Rushton, M. T., Bewsher, D., Walter, F. M., Eyres, S. P. S., Hounsell, R., and Darnley, M. J. 2013, *Mon. Not. Roy. Astron. Soc.*, **438**, 3483.

Inno, L., *et al.* 2015, *Astron. Astrophys.*, **576A**, 30.

Jacoby, G. H., and Pierce, M. J. 1996, *Astron. J.*, **112**, 723.

Kafka, S. 2017, variable star observations from the AAVSO International Database (https://www.aavso.org/aavso-international-database-aid).

Leibowitz, E. M., and Formiggini, L. 2015, *Astron. J.*, **150**, 52.

Mayangsari, L., Priyatikanto, R., and Putra, M. 2014, AIP Conf. Proc., 1589, 37.

Pierce, M. J., and Jacoby, G. H. 1995, *Astron. J.*, **110**, 2885.

Price, A., Foster, G., and Skiff, B. 2007, *The Precision of Visual Estimates of Variable Stars* (AAS Meeting, January 2007).[1]

Schaefer, B. J., 1996, *Astron. J.*, **111**, 1668.

Simonsen, M. A. 2004, *J. Amer. Assoc. Var. Star Obs.*, **33**, 65.

Stanton, R. H. 1999, *J. Amer. Assoc. Var. Star Obs.*, **27**, 97.

Surina, F., Hounsell, R. A., Bode, M. F., Darnley, M. J., Harman, D. J., and Walter, F. M. 2014, in *Stella Novae: Past and Future Decades*, eds. P. A. Woudt, V. A. R. M. Ribiero, ASP Conf. Ser. 490, 169.

Trumpler, R. J., and Weaver, H. F. 1953, *Statistical Astronomy*, Univ. California Press (Dover edition 1962), Dover Publ., New York, 157.

Whiting, A. B. 2012, *Observatory*, **132**, 148.

Willett, K. W., *et al.* 2013, *Mon. Not. Roy. Astron. Soc.*, **433**, 2835.

Williams, D. B. 1987, *J. Amer. Assoc. Var. Star Obs.*, **16**, 118.

Zissell, R. E. 2003, *J. Amer. Assoc. Var. Star Obs.*, **31**, 128.

---

[1] Price, Foster, G., and Skiff (2007), https://www.aavso.org/sites/default/files/publications/staff_pubs/price_visual_precision.pdf.

**Appendix A: Information on the stars in this study.**

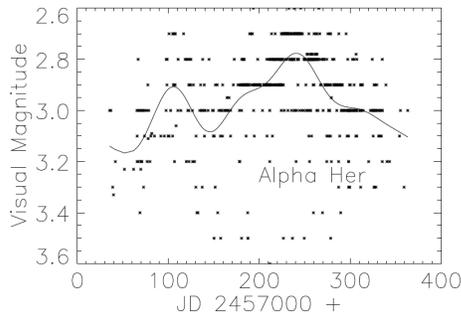

Figure A1a. Visual observations of α Herculis, with the best-fit average light curve superimposed.

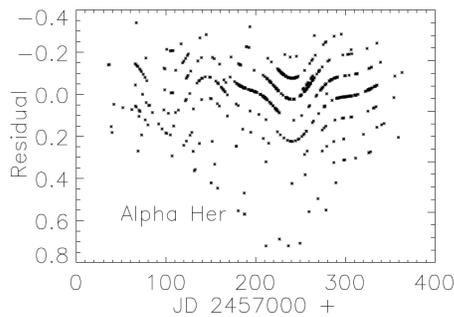

Figure A1b. Residuals of the observations of α Herculis, with the best-fit average light curve subtracted. Most (but not all) visual observers report to the nearest tenth-magnitude, leading to some artifacts.

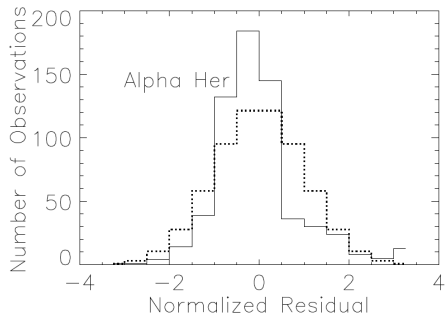

Figure A1c. Normalized residuals of α Herculis observations (solid line), with a Gaussian distribution superimposed (dotted line). The residuals are clearly non-Gaussian.

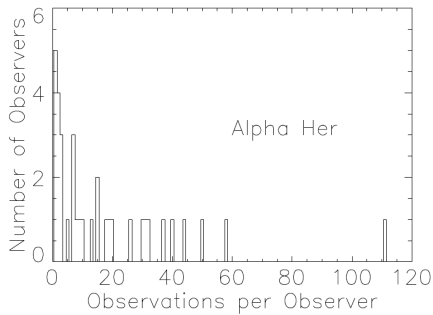

Figure A1d. The distribution of activity among observers of α Herculis. The data are not dominated by any single observer, nor by the single-digit observers.

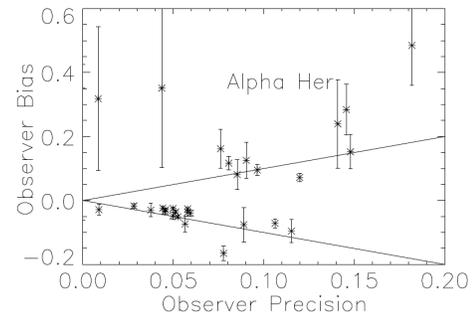

Figure A1e. Comparison of each observer's average distance from the best-fit curve (bias) with precision (the standard deviation about that average). Error bars in bias are derived by dividing the bias value by the square root of the number of observations; they are not shown for precision in order to keep the plot readable. Bias is smaller than precision for observers inside the lines drawn. Note that for most observers outside the lines, bias is not well-determined owing to few observations. Most observations fall inside the funnel.

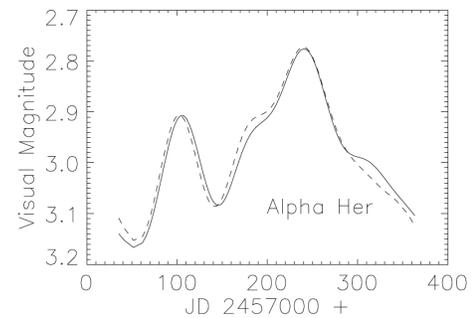

Figure A1f. The original light curve of α Herculis (solid curve) compared with the reconstructed one (dashed). The reconstructed curve was determined by imposing Gaussian residuals about the original curve, and redoing the fit.

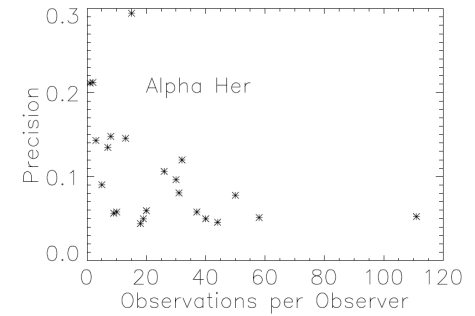

Figure A1g. The variation of observer precision with number of observations. There is some tendency for low-activity observers to have worse precision, but it depends mostly on a few outlying points.

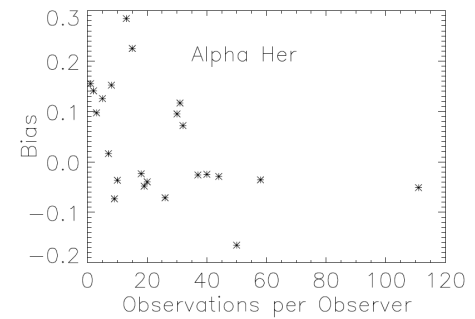

Figure A1h. The variation of observer bias with number of observations. There is no significant tendency for bias to become smaller with more activity.

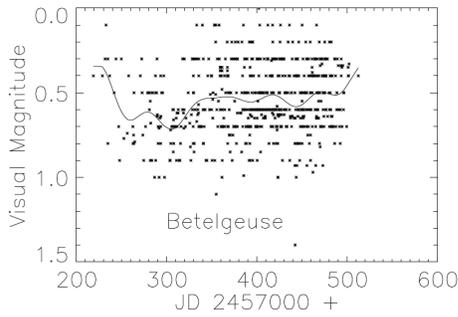

Figure A2a. Visual observations of Betelgeuse, α Ori, with the best-fit average light curve superimposed.

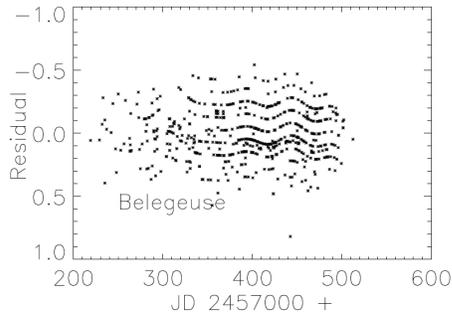

Figure A2b. Residuals of the observations of Betelgeuse, with the best-fit average light curve subtracted. Most (but not all) visual observers report to the nearest tenth-magnitude, leading to some artifacts.

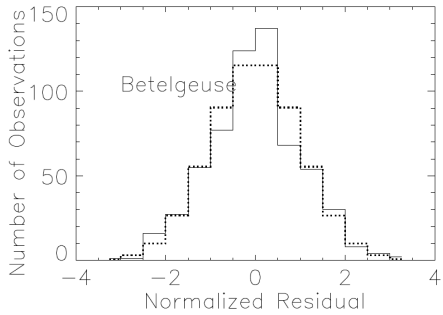

Figure A2c. Normalized residuals of Betelgeuse observations (solid line), with a Gaussian distribution superimposed (dotted line). The residuals are clearly non-Gaussian.

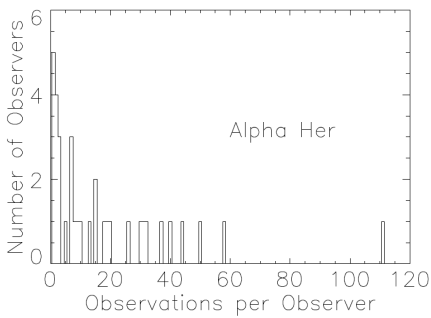

Figure A2d. The distribution of activity among observers of Betelgeuse. The data are not dominated by any single observer, nor by the single-digit observers.

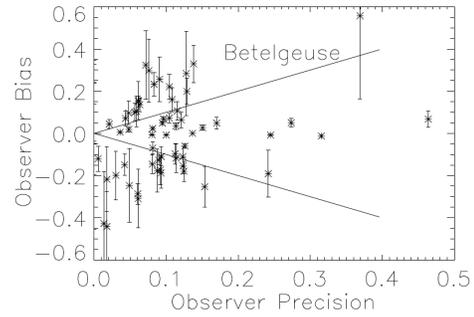

Figure A2e. Comparison of each observer's average distance from the best-fit curve (bias) with precision (the standard deviation about that average). Error bars in bias are derived by dividing the bias value by the square root of the number of observations; they are not shown for precision in order to keep the plot readable. Bias is smaller than precision for observers inside the lines drawn. Note that for many observers outside the lines, bias is not well-determined owing to few observations. Most observations fall inside the funnel.

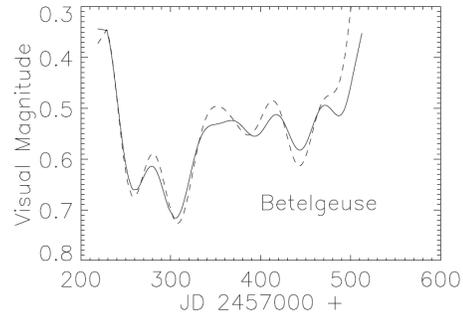

Figure A2f. The original light curve of Betelgeuse (solid curve) compared with the reconstructed one (dashed). The reconstructed curve was determined by imposing Gaussian residuals about the original curve, and redoing the fit.

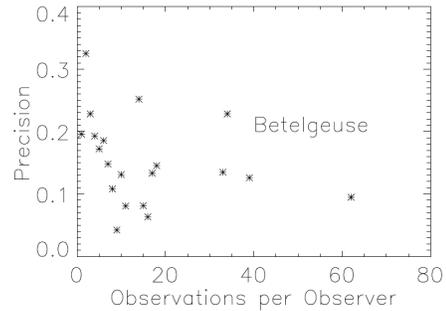

Figure A2g. The variation of observer precision with number of observations. There is no significant trend.

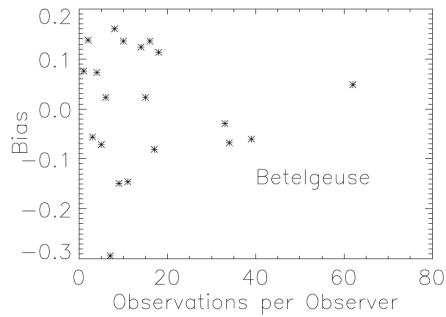

Figure A2h. The variation of observer bias with number of observations. There are too few points beyond 20 observations to reach a conclusion on overall trend; up to that point, there is certainly none.

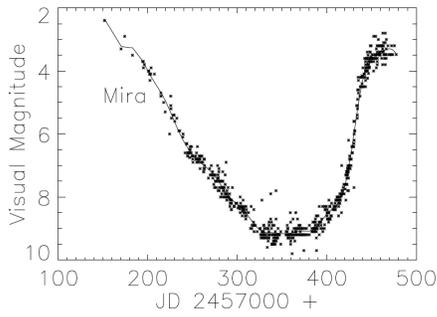

Figure A3a. Visual observations of Mira, o Ceti, with the best-fit average light curve superimposed.

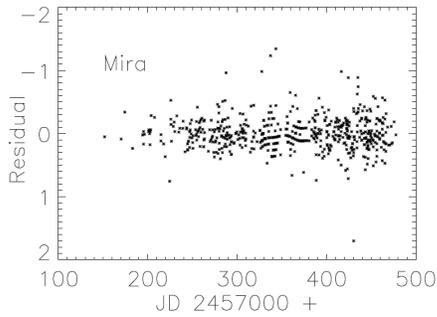

Figure A3b. Residuals of observations of Mira, with the best-fit light curve subtracted. The practice of most (but not all) observers of reporting to the nearest tenth-magnitude leads to some aritfacts.

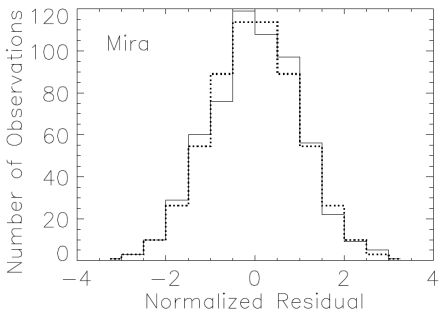

Figure A3c. Normalized residuals of Mira observations (solid line), with a Gaussian distribution superimposed (dotted line). The residuals are significantly non-Gaussian.

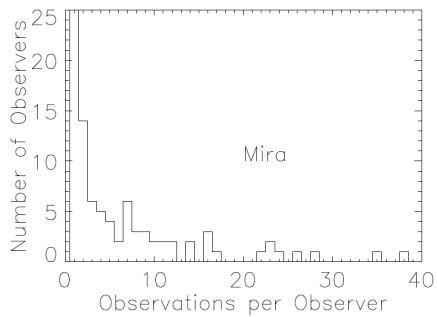

Figure A3d. The distribution of activity among observers of Mira. The data are not dominated by any single observer, nor by the single-digit observers.

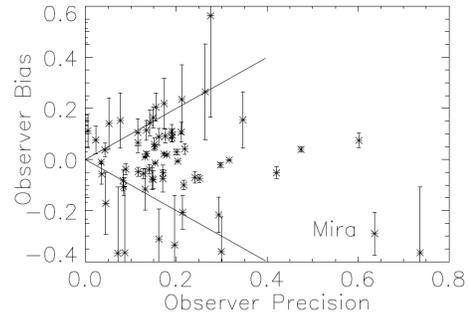

Figure A3e. Comparison of each observer's average distance from the best-fit curve (bias) with precision (the standard deviation about that average). Error bars in bias are derived by dividing the bias value by the square root of the number of observations; they are not shown for precision in order to keep the plot readable. Bias is smaller than precision for observers inside the lines drawn. Note that for most observers outside the lines, bias is not well-determined owing to few observations. Most observations fall inside the funnel.

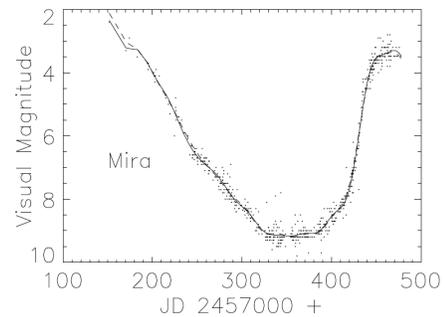

Figure A3f. Mira's original light curve (solid curve) compared with the reconstructed one (dashed). The reconstructed curve was determined by imposing Gaussian residuals about the original curve, and redoing the fit.

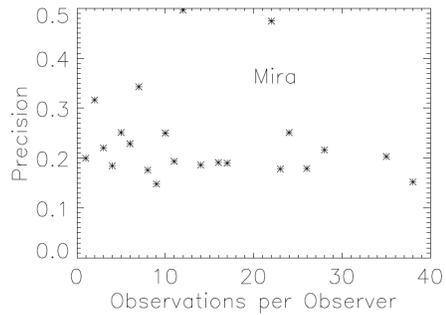

Figure A3g. The variation of observer precision with number of observations. There is no apparent trend.

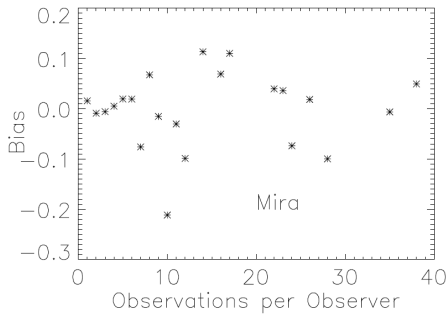

Figure A3h. The variation of observer bias with number of observations. Again, there is no apparent trend.

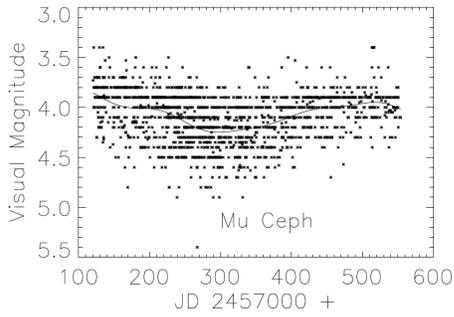

Figure A4a. Visual observations of μ Cephei, with the best-fit average light curve superimposed.

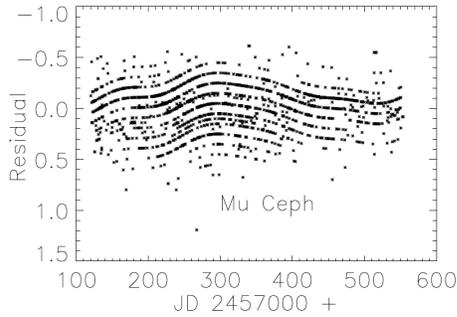

Figure A4b. Residuals of the observations of μ Cephei, with the best-fit average light curve subtracted. Most (but not all) visual observers report to the nearest tenth-magnitude, leading to some artifacts.

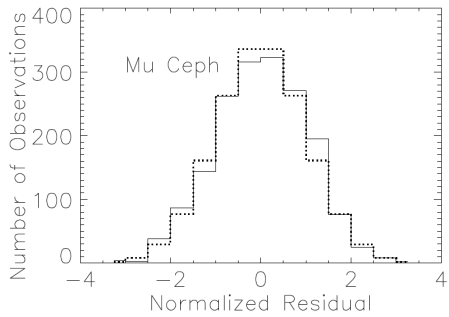

Figure A4c. Normalized residuals of μ Cephei observations (solid line), with a Gaussian distribution superimposed (dotted line). While similar, the residuals are significantly non-Gaussian.

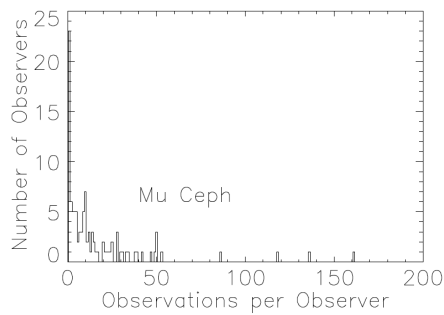

Figure A4d. The distribution of activity among observers of μ Cephei. The data are not dominated by any single observer, nor by the single-digit observers.

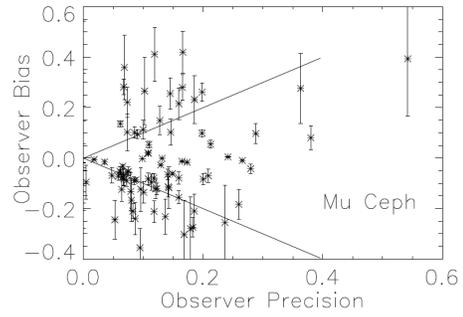

Figure A4e. Comparison of each observer's average distance from the best-fit curve (bias) with precision (the standard deviation about that average). Error bars in bias are derived by dividing the bias value by the square root of the number of observations; they are not shown for precision in order to keep the plot readable. Bias is smaller than precision for observers inside the lines drawn. Note that for many observers outside the lines, bias is not well-determined owing to few observations. Most observations fall inside the funnel.

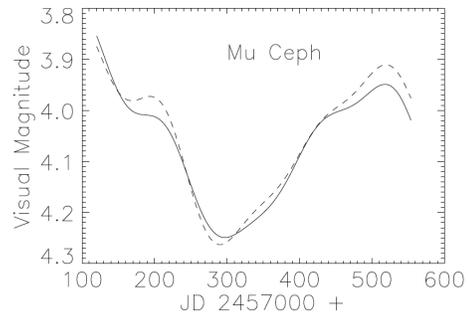

Figure A4f. The original light curve of μ Cephei (solid curve) compared with the reconstructed one (dashed). The reconstructed curve was determined by imposing Gaussian residuals about the original curve, and redoing the fit.

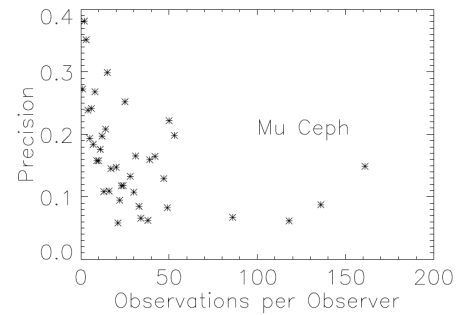

Figure A4g. The variation of observer precision with number of observations. There is no trend visible in the region below about 60 observations; above that, there are too few points to reach any conclusion.

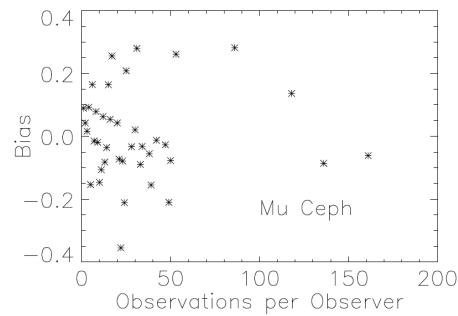

Figure A4h. The variation of observer bias with number of observations. There is no significant tendency for bias to become smaller with more activity.

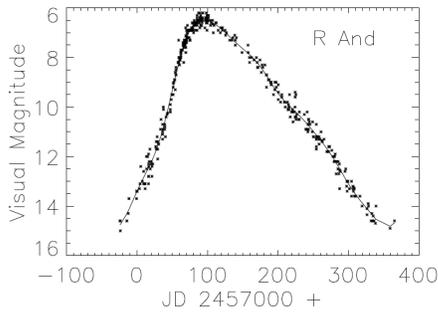

Figure A5a. Visual observations of R Andromedae, with the best-fit average light curve superimposed.

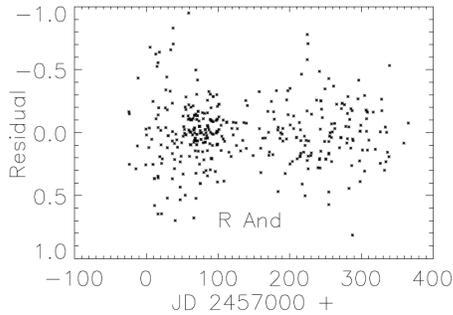

Figure A5b. Residuals of the observations of R Andromedae, with the best-fit average light curve subtracted. There is an apparent concentration of observations around maximum.

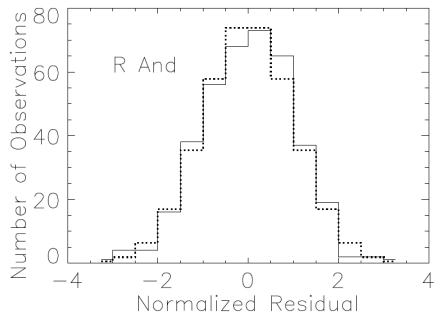

Figure A5c. Normalized residuals of R Andromedae observations (solid line), with a Gaussian distribution superimposed (dotted line). The residuals are significantly non-Gaussian.

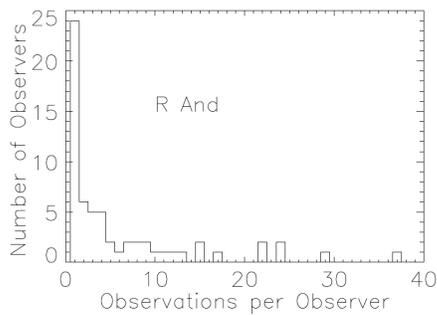

Figure A5d. The distribution of activity among observers of R Andromedae. The data are not dominated by any single observer, nor by the single-digit observers.

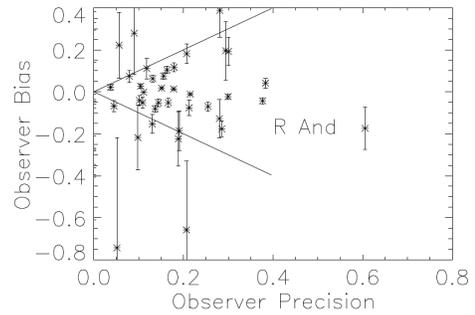

Figure A5e. Comparison of each observer's average distance from the best-fit curve (bias) with precision (the standard deviation about that average). Error bars in bias are derived by dividing the bias value by the square root of the number of observations; they are not shown for precision in order to keep the plot readable. Bias is smaller than precision for observers inside the lines drawn. Note that for most observers outside the lines, bias is not well-determined owing to few observations. Most observations fall inside the funnel.

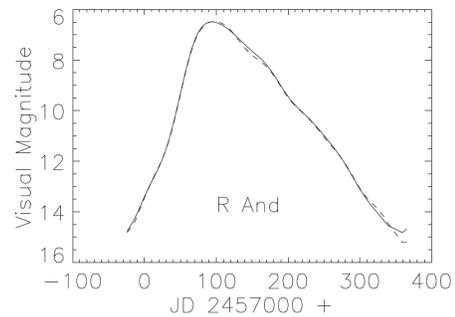

Figure A5f. The original light curve of R Andromedae (solid curve) compared with the reconstructed one (dashed). The reconstructed curve was determined by imposing Gaussian residuals about the original curve, and redoing the fit.

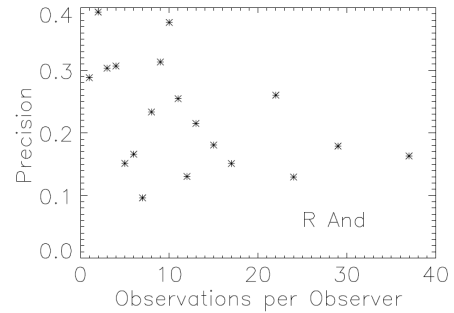

Figure A5g. The variation of observer precision with number of observations. There is some tendency for low-activity observers to have worse precision, but it depends mostly on a few outlying points.

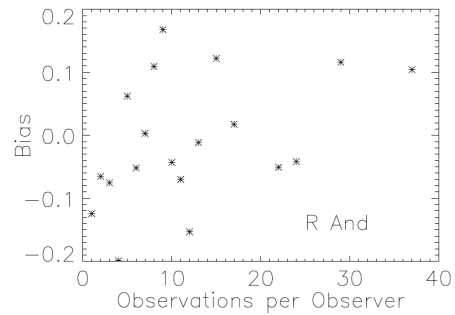

Figure A5h. The variation of observer bias with number of observations. There is no significant tendency for bias to become smaller with more activity.

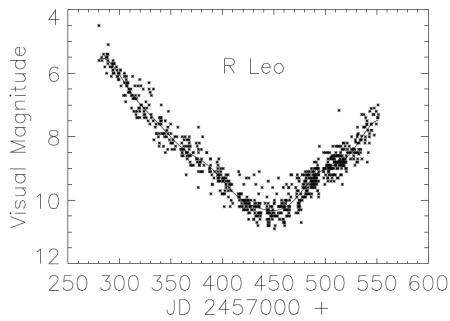

Figure A6a. Visual observations of R Leonis, with the best-fit average light curve superimposed.

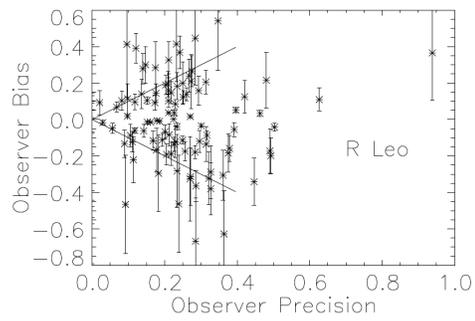

Figure A6e. Comparison of each observer's average distance from the best-fit curve (bias) with precision (the standard deviation about that average). Error bars in bias are derived by dividing the bias value by the square root of the number of observations; they are not shown for precision in order to keep the plot readable. Bias is smaller than precision for observers inside the lines drawn. Note that for most observers outside the lines, bias is not well-determined owing to few observations. Most observations fall inside the funnel.

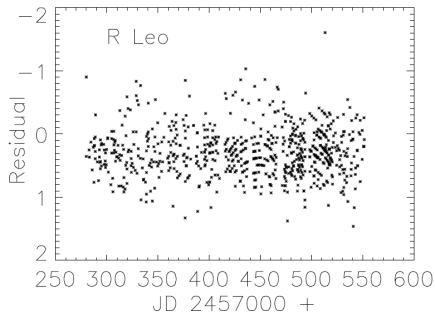

Figure A6b. Residuals of the observations of R Leonis, with the best-fit average light curve subtracted. Most (but not all) visual observers report to the nearest tenth-magnitude, leading to some artifacts.

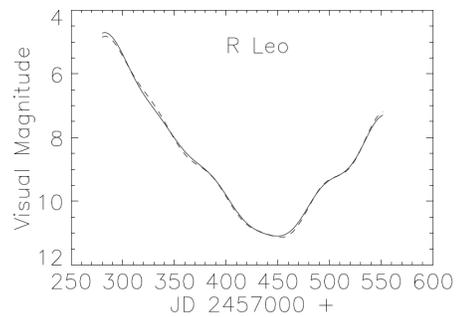

Figure A6f. The original light curve of R Leonis (solid curve) compared with the reconstructed one (dashed). The reconstructed curve was determined by imposing Gaussian residuals about the original curve, and redoing the fit.

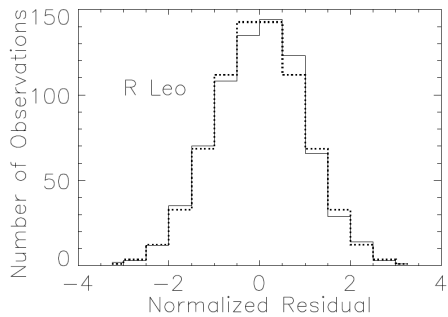

Figure A6c. Normalized residuals of R Leonis observations (solid line), with a Gaussian distribution superimposed (dotted line). The residuals are significantly non-Gaussian.

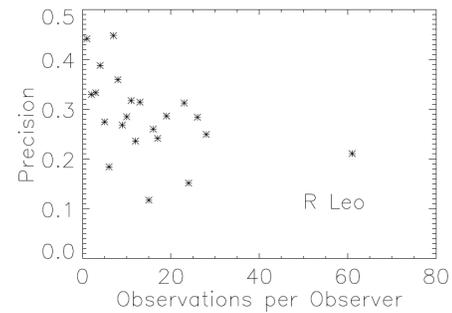

Figure A6g. The variation of observer precision with number of observations. There is a weak tendency for low-activity observers to have worse precision, but it depends mostly on a few outlying points.

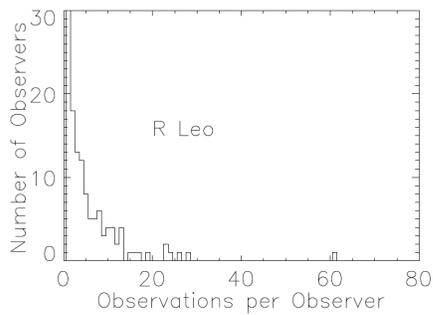

Figure A6d. The distribution of activity among observers of R Leonis. The data are not dominated by any single observer, nor by the single-digit observers.

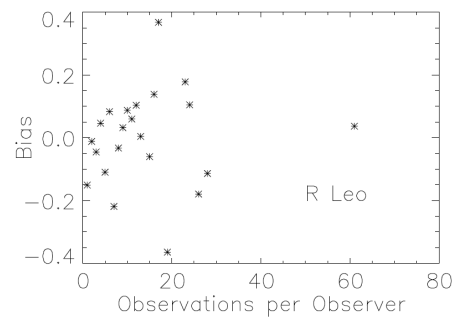

Figure A6h. The variation of observer bias with number of observations. There is no significant tendency for bias to become smaller with more activity.

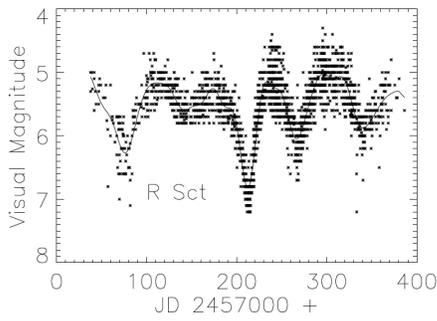

Figure A7a. Visual observations of R Scuti, with the best-fit average light curve superimposed.

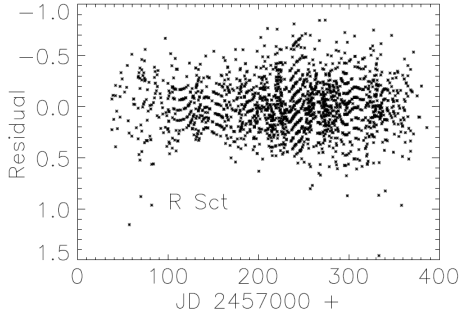

Figure A7b. Residuals of the observations of R Scuti, with the best-fit average light curve subtracted. Most (but not all) visual observers report to the nearest tenth-magnitude, leading to some artifacts.

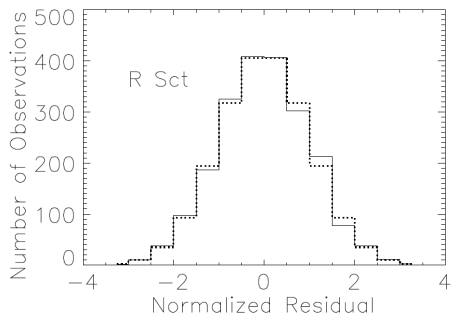

Figure A7c. Normalized residuals of R Scuti observations (solid line), with a Gaussian distribution superimposed (dotted line). The residuals are significantly non-Gaussian.

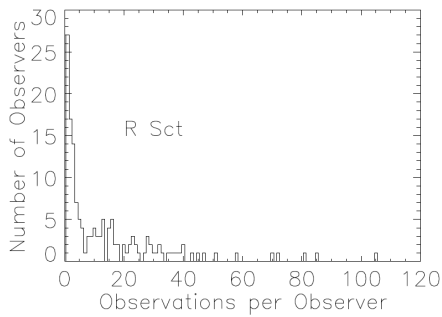

Figure A7d. The distribution of activity among observers of R Scuti. The data are not dominated by any single observer, nor by the single-digit observers.

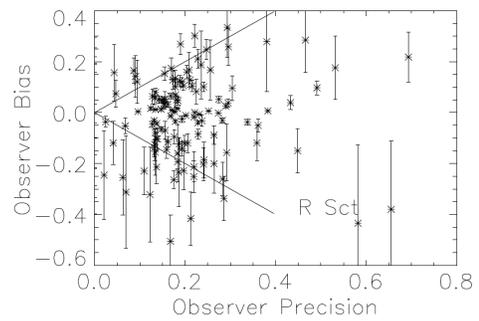

Figure A7e. Comparison of each observer's average distance from the best-fit curve (bias) with precision (the standard deviation about that average). Error bars in bias are derived by dividing the bias value by the square root of the number of observations; they are not shown for precision in order to keep the plot readable. Bias is smaller than precision for observers inside the lines drawn. Note that for most observers outside the lines, bias is not well-determined owing to few observations. Most observations fall inside the funnel.

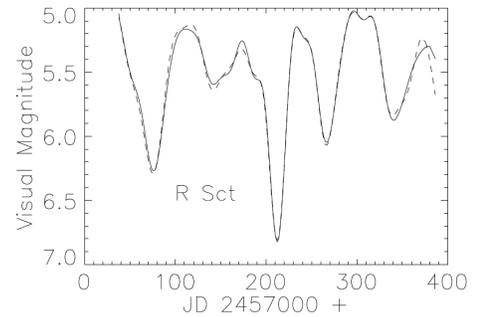

Figure A7f. The original light curve of R Scuti (solid curve) compared with the reconstructed one (dashed). The reconstructed curve was determined by imposing Gaussian residuals about the original curve, and redoing the fit.

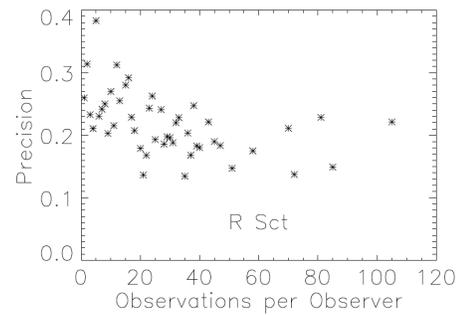

Figure A7g. The variation of observer precision with number of observations. There is some tendency for low-activity observers to have worse precision, but it depends mostly on a few outlying points.

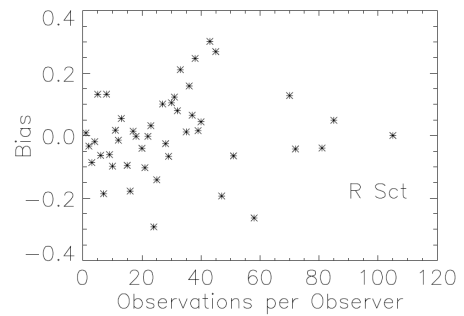

Figure A7h. The variation of observer bias with number of observations. There is no significant tendency for bias to become smaller with more activity up to about 60 observations; beyond that, there are too few points to allow a conclusion.

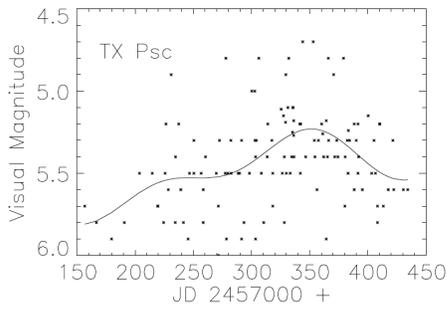

Figure A8a. Visual observations of TX Piscium, with the best-fit average light curve superimposed.

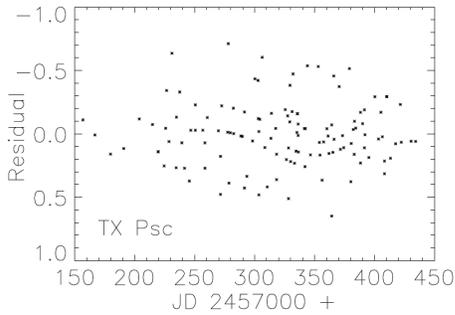

Figure A8b. Residuals of the observations of TX Piscium, with the best-fit average light curve subtracted. Most (but not all) visual observers report to the nearest tenth-magnitude, leading to some artifacts.

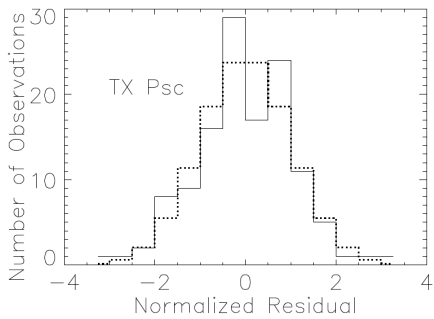

Figure A8c. Normalized residuals of TX Piscium observations (solid line), with a Gaussian distribution superimposed (dotted line). The residuals are consistent with being Gaussian.

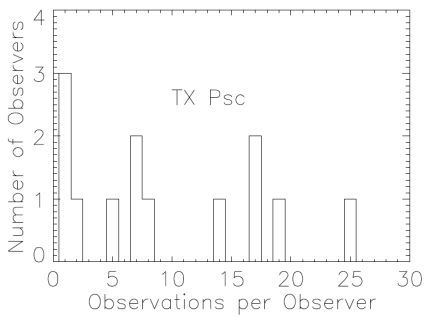

Figure A8d. The distribution of activity among observers of TX Piscium. There is a smaller proportion of low-activity observers compared with the other stars in this study.

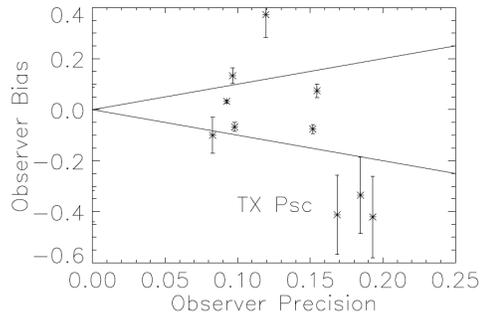

Figure A8e. Comparison of each observer's average distance from the best-fit curve (bias) with precision (the standard deviation about that average). Error bars in bias are derived by dividing the bias value by the square root of the number of observations; they are not shown for precision in order to keep the plot readable. Bias is smaller than precision for observers inside the lines drawn. Note that for most observers outside the lines, bias is not well-determined owing to few observations. Most observations fall inside the funnel.

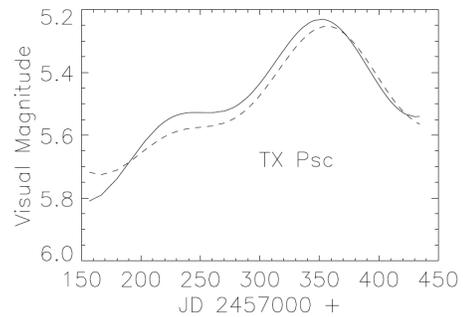

Figure A8f. The original light curve of TX Piscium (solid curve) compared with the reconstructed one (dashed). The reconstructed curve was determined by imposing Gaussian residuals about the original curve, and redoing the fit.

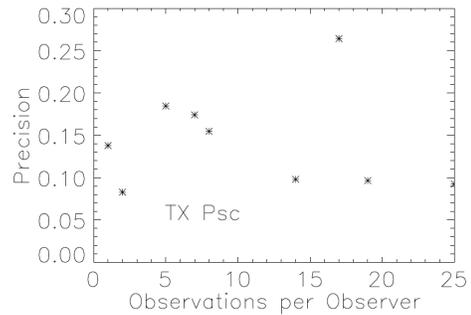

Figure A8g. The variation of observer precision with number of observations. There is no apparent trend.

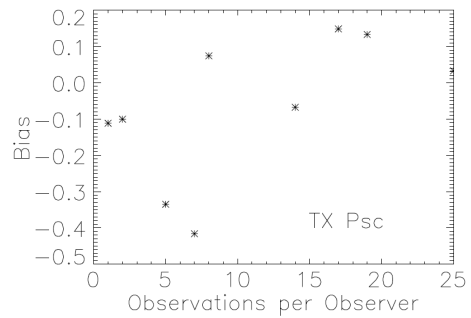

Figure A8h. The variation of observer bias with number of observations. There is no significant tendency for bias to become smaller with more activity.

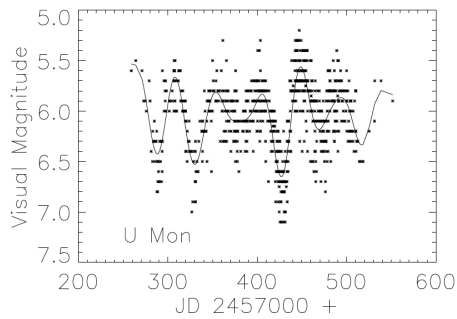

Figure A9a. Visual observations of U Monocerotis, with the best-fit average light curve superimposed.

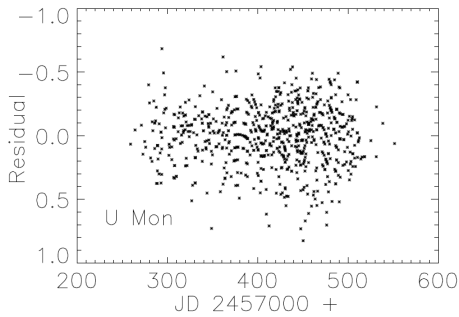

Figure A9b. Residuals of the observations of U Monocerotis, with the best-fit average light curve subtracted. Most (but not all) visual observers report to the nearest tenth-magnitude, leading to some artifacts.

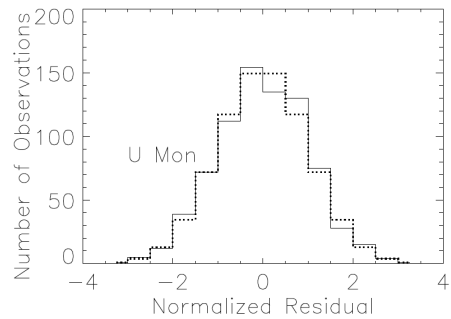

Figure A9c. Normalized residuals of U Monocerotis observations (solid line), with a Gaussian distribution superimposed (dotted line). The residuals are significantly non-Gaussian.

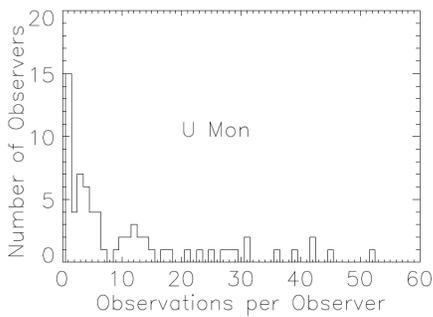

Figure A9d. The distribution of activity among observers of U Monocerotis. The data are not dominated by any single observer, nor by the single-digit observers.

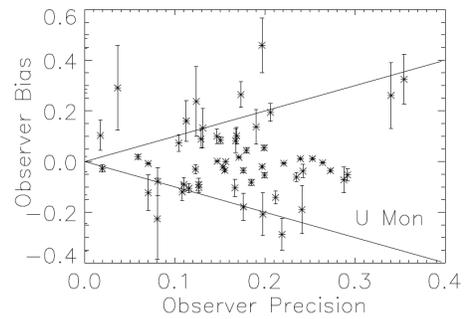

Figure A9e. Comparison of each observer's average distance from the best-fit curve (bias) with precision (the standard deviation about that average). Error bars in bias are derived by dividing the bias value by the square root of the number of observations; they are not shown for precision in order to keep the plot readable. Bias is smaller than precision for observers inside the lines drawn. Note that for most observers outside the lines, bias is not well-determined owing to few observations. Most observations fall inside the funnel.

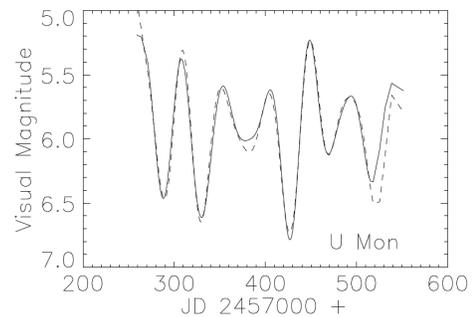

Figure A9f. The original light curve of U Monocertis (solid curve) compared with the reconstructed one (dashed). The reconstructed curve was determined by imposing Gaussian residuals about the original curve, and redoing the fit.

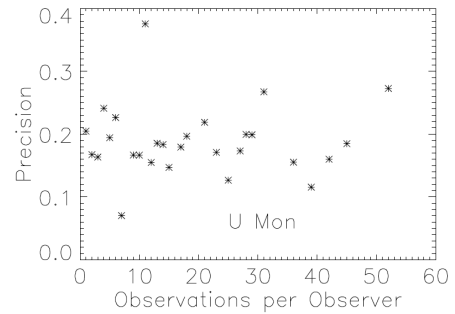

Figure A9g. The variation of observer precision with number of observations. There is no apparent trend.

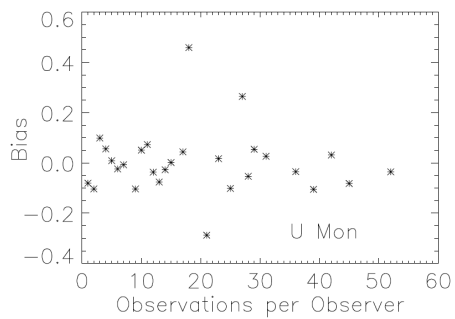

Figure A9h. The variation of observer bias with number of observations. There is no significant tendency for bias to become smaller with more activity.

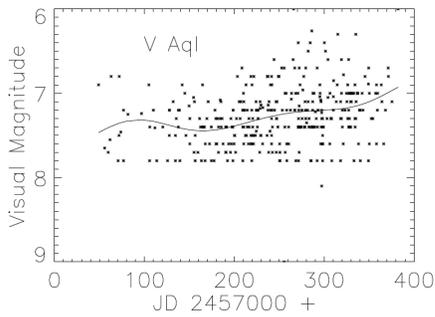

Figure A10a. Visual observations of V Aquilae, with the best-fit average light curve superimposed.

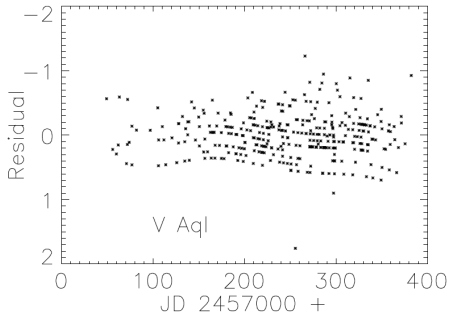

Figure A10b. Residuals of the observations of V Aquilae, with the best-fit average light curve subtracted. Most (but not all) visual observers report to the nearest tenth-magnitude, leading to some artifacts.

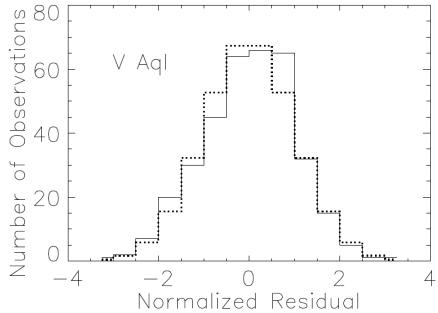

Figure A10c. Normalized residuals of V Aquilae observations (solid line), with a Gaussian distribution superimposed (dotted line). The residuals are significantly non-Gaussian.

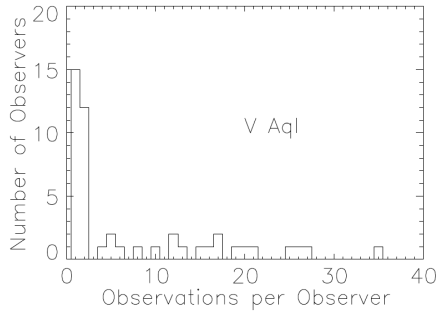

Figure A10d. The distribution of activity among observers of V Aquilae. The data are not dominated by any single observer, nor by the single-digit observers.

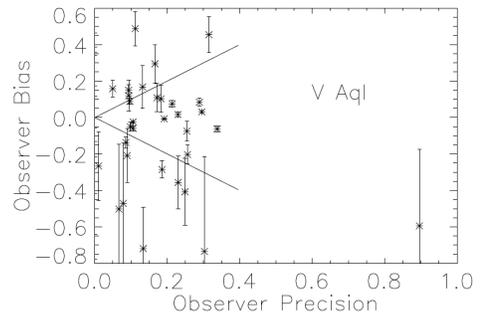

Figure A10e. Comparison of each observer's average distance from the best-fit curve (bias) with precision (the standard deviation about that average). Error bars in bias are derived by dividing the bias value by the square root of the number of observations; they are not shown for precision in order to keep the plot readable. Bias is smaller than precision for observers inside the lines drawn. Note that for most observers outside the lines, bias is not well-determined owing to few observations. Most observations fall inside the funnel.

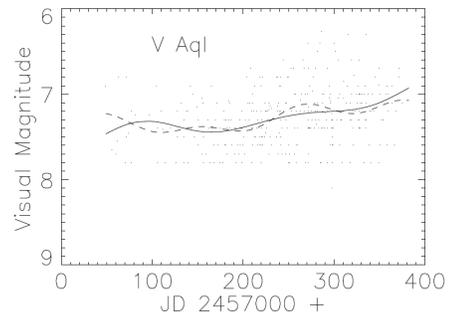

Figure A10f. The original light curve of V Aquilae (solid curve) compared with the reconstructed one (dashed). The reconstructed curve was determined by imposing Gaussian residuals about the original curve, and redoing the fit.

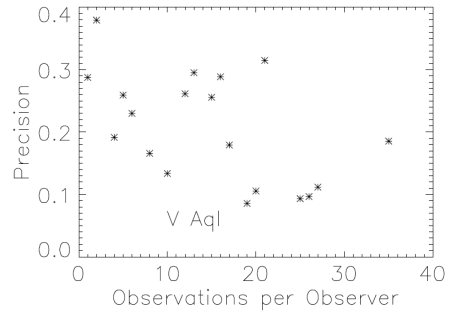

Figure A10g. The variation of observer precision with number of observations. There is no clear trend.

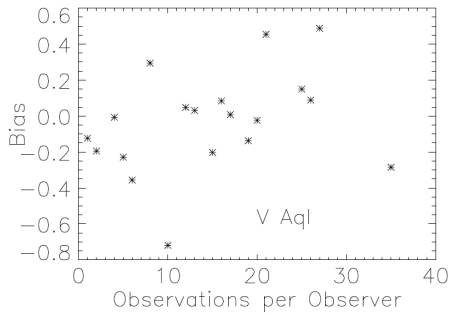

Figure A10h. The variation of observer bias with number of observations. There is no significant tendency for bias to become smaller with more activity.